\documentclass[apjl]{emulateapj}

\shorttitle{Barium production in low mass stars}
\shortauthors{D'Orazi et al.}

\begin{document}


\title{Enhanced production of barium in low-mass stars: evidence from open
clusters}


\author{Valentina D'Orazi}
\affil{Dipartimento di Astronomia e Scienza dello Spazio, Universit\`a
di Firenze, Firenze, Italy}
\affil{INAF --Osservatorio Astrofisico di Arcetri, Firenze, Italy}
\email{vdorazi@arcetri.astro.it}
\author{Laura Magrini}
\author{Sofia Randich}
\author{Daniele Galli}
\affil{INAF --Osservatorio Astrofisico di Arcetri, Firenze, Italy}
\author{Maurizio Busso}
\affil{Dipartimento di Fisica, Universit\`a di Perugia, and Sezione INFN, Perugia, Italy}
\author{Paola Sestito}
\affil{INAF --Osservatorio Astrofisico di Arcetri, Firenze, Italy}


\begin{abstract}
We report the discovery of a trend of increasing barium abundance with
decreasing age for a large sample of Galactic open clusters. The observed
pattern of [Ba/Fe] vs. age can be reproduced with a
Galactic chemical evolution model only assuming a higher Ba yield
from the $s$-process in low-mass stars than the average one suggested
by parametrized models of neutron-capture nucleosynthesis. We show
that this is possible in a scenario where the efficiency of
the extra-mixing processes producing the neutron source $^{13}$C is
anti-correlated with the initial mass, with a larger efficiency
for lower masses. This is similar to the known trend of extended mixing
episodes acting in H-rich layers and might suggest a common physical
mechanism.
\end{abstract}

\keywords{Galaxy: disk, abundances, evolution; open clusters and
associations: general, abundances}

\section{Introduction}
\label{sec:intro}

Barium is one of the so-called heavy neutron-capture elements. It is
formed by neutron captures occurring through two major mechanisms:
({\em i}\/) the slow capture process ($s$-process) and ({\em ii}\/) the
rapid one ($r$-process).  The $r$-process is thought to occur in
high-mass stars, $M \gtrsim 8~M_\odot$, exploding as type-II SNe
(e.g., Kratz et
al. 2007; Travaglio et al.~1999). The $s$-process instead comes from low- and
intermediate-mass stars, $1~M_\odot\lesssim M\lesssim 3~M_\odot$ (e.g.,
Busso et al.~2001), thanks to the repeated neutron expositions
guaranteed by the activation of the neutron sources
$^{13}$C($\alpha$,n)$^{16}$O and $^{22}$Ne($\alpha$,n)$^{25}$Mg. The
$s$-process contribution from low-mass stars had a dominant role in the
solar mixture of Ba nuclei, so that in the Sun Ba comes from the
s-process at about the 80\% level (Arlandini et al. 1999).

Barium abundances have been measured in stars with different
metallicities (e.g., Burris et al.~2000 for halo stars;
Cescutti et al.~2006 and Bensby et al.~2005 for thin and thick disk
stars), constraining the nucleosynthetic processes at the origin of
Ba.  However, several details of the Ba production are still a matter
of debate due to the free parameters affecting the models (see,
e.g., for the $r$-process Mathews et al.~1992; Pagel \&
Tautvai\v{s}ien\.{e}~1997; Travaglio et al.~1999; for the $s$-process
Busso et al. 2001). In the $s$-process, the main source of uncertainty
is the still unknown physical mechanism promoting the penetration of
protons into the He intershell zone at dredge-up; this penetration is
necessary for explaining the formation of the main neutron source
$^{13}$C.

In this {\em Letter} we address the evolution of Ba with age 
by means of Ba measurements in a sample of open
clusters (OCs) and by comparing the empirical pattern
with the results of a Galactic chemical evolution model.
We find an unexpected trend, in
apparent conflict with the currently adopted average Ba yields. We also
identify in which direction the average yields should be corrected and
why this is physically justified.

\section{Sample and analysis}
\label{sec:sample}

Our sample (see Table~1) consists of 20 OCs spanning a wide interval in
age ($\sim$30~Myr--8~Gyr), Galactocentric radius (7--22~kpc), and
metallicity ($-0.3\leq \mbox{[Fe/H]}\leq +0.4$). We analyzed a large
collection of high-resolution spectra obtained by our group during the years
with different
instruments: UVES in slit-mode and FLAMES/UVES on the ESO VLT, CASPEC
on the ESO 3.6m telescope, the Red-Long camera echelle spectrograph at
CTIO, and SARG on the Italian National Telescope Galileo.  
The spectra of unevolved members were analyzed in
the first 10 OCs in Table 1, while the analysis of the older and more
distant clusters is based on giant stars.

Barium abundances were computed by means of equivalent width (EW)
analysis and by using the driver {\it blends} in MOOG (Sneden,
1973 --2002 version). The grid of 1-D LTE model atmospheres of Kurucz
(1993) was employed.  Our analysis uses two Ba~lines:
$\lambda\lambda$ 5853 and 6496~\AA.  As well known, Ba lines are
affected by hyperfine-splitting (hfs); thus we included in our line
list the hyperfine structure and isotopic splitting, following the
approach by McWilliam~(1998) and using $\log~gf$ values published
there.  An 81\% $s$-process composition was adopted for the Sun and the
sample stars. Stark and radiative broadening were treated in the
standard way and for collisional damping we adopted the classical
Uns\"old (1955) approximation.

As a first step, we derived the solar Ba abundance using the same
method as for the sample stars: for the Sun we adopted T$_{\rm eff}=5770$~K,
$\log$~g=4.44 and $\xi$=1.1~kms/s (Randich et al.~2006), obtaining $\log$
n(Ba)$=2.22$ and $\log$n(Ba)$=2.26$ for $\lambda=5853$ and
6496~\AA, respectively. The average solar value $\log$
n(Ba)$=2.24\pm 0.02$ agrees very well with the meteoritic
abundance of $\log$n(Ba)$=$2.22 (Grevesse et al.~1996). Our
analysis is strictly differential with respect to the Sun: we applied a
line by line solar correction, canceling out the uncertainties on the
adopted set of atomic parameters ($\log gf$).  The feature at
$\lambda=6496$~\AA~always yields higher abundances, suggesting,
as also stressed by Prochaska et al.~\cite{prochaska00}, the presence
of an unidentified blend; however, the differential nature of our
analysis should minimize this effect, in particular for dwarf stars
with close-to-solar stellar parameters. Ba abundances were obtained adopting
stellar parameters and uncertainties derived in our previous studies.

Errors in Ba abundances for each star are due to uncertainties in EWs
and stellar parameters. The standard deviation from the mean of each
cluster instead provides a good estimate of the global errors on the
average abundances\footnote{The line list with the EWs measurements as
well as stellar parameters, and Ba abundances of each single star along
with errors are available upon request.}. In order to estimate
systematic errors, we compared our results with measurements from the
literature for six clusters. We found a good agreement within the
uncertainties ($\Delta$[Ba/Fe]$<$0.06dex) for M67, Berkeley~20, and
NGC~6253; the differences are instead larger for the Hyades
($\Delta$[Ba/Fe]$=-0.2$~dex), Collinder~261
($\Delta$[Ba/Fe]=$+0.2$~dex), and Berkeley~29
($\Delta$[Ba/Fe]$=+0.14$~dex).  These results show that, whereas
discrepancies with the literature are present, these are not
systematic; in particular our Ba abundances are not systematically
higher. Also, given the old age of Cr 261 and Be~29, our higher [Ba/Fe]
ratio for these clusters does not affect our final conclusions.

\section{Results}
\label{sec:disc}

The results are summarized in Table~\ref{results}, where we report the
average [Ba/Fe] ratios for each cluster along with the standard
deviation from the mean. In Figure~1 we show [Ba/Fe]
ratios as a function of cluster age; different symbols indicate dwarf
and giant members. In the figure we also show the solar [Ba/Fe] as well
as our own measurement for $\alpha$ Centauri performed by using EWs
published by Porto de Mello et al.~\cite{por08}. Our Ba abundance is
in excellent agreement with this previous result
($\Delta$[Ba/Fe]$<0.02$~dex). The trend of increasing [Ba/Fe] from
the oldest clusters to the youngest ones is evident. The figure also
shows that [Ba/Fe] ratios are systematically higher for clusters where
giants have been analyzed.  We suggest that this might be due to the
fact that our analysis is not strictly differential for giant stars, as
their parameters are different from the solar ones. \\
First, the question arises whether the observed trend of increasing
[Ba/Fe] with decreasing age is real or rather due to any effect that,
for a given intrinsic Ba abundance, would increase the observed EWs. In
particular, members of the youngest clusters in our sample (ages below
$\sim 500$~Myr) are characterized by strong magnetic fields and are
chromospherically very active; thus Ba~II line strengths might be
affected by both a ``magnetic intensification mechanism" (Landi
degl'Innocenti \& Landolfi 2004) and by non-LTE (NLTE) effects.  In order
to check whether magnetic intensification could be a problem, we
measured the EWs of the weak Eu~{\rm II} 6437~\AA~feature in the sample
stars.  Magnetic intensification would cause a similar (if not
larger) effect for Eu, whose hyper-fine structure is notoriously more
complex (the number of Zeeman sub-levels is higher). We found that, at
variance with Ba, Eu EWs do not increase with the age, thus excluding
that this process affects Ba EWs.

On the other hand, NLTE effects might be important for Ba~II features,
in particular, for the $\lambda 6496$ \AA~line and for stars with
close-to-solar metallicity (Mashonkina et al. 1999). It is very likely
that NLTE effects are amplified in young stars due to the presence of
hot chromospheres.  While we cannot quantify this effect, we believe
that [Ba/Fe] for clusters younger than $\sim 500$~Myr are likely
overestimated.

\begin{figure}
\includegraphics[width=9cm]{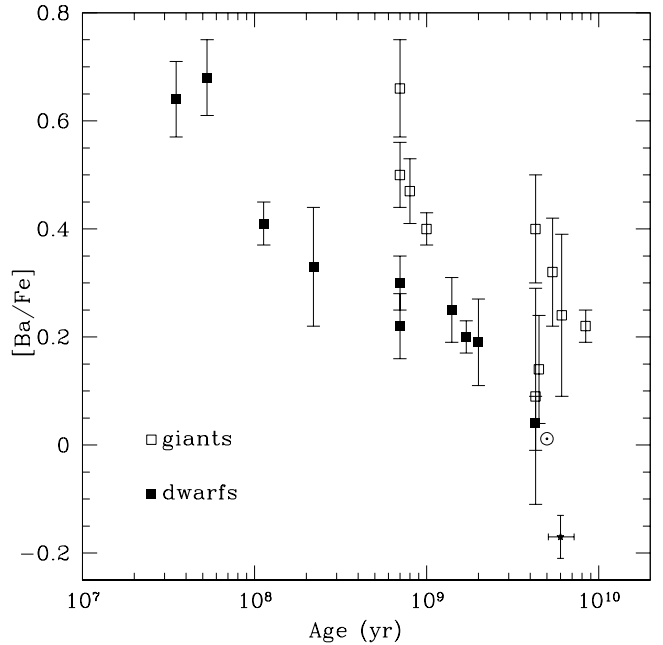}
\caption{Average [Ba/Fe] as function of age for the sample clusters; 
filled and open  squares denote clusters whose analysis is based on dwarf
and giant stars, respectively. We also show
the solar value (solar symbol) and that of the
$\alpha$~Centauri (star with errorbars). 
}
\label{bafe}
\end{figure}

\begin{figure}
\includegraphics[height=9cm, angle=270]{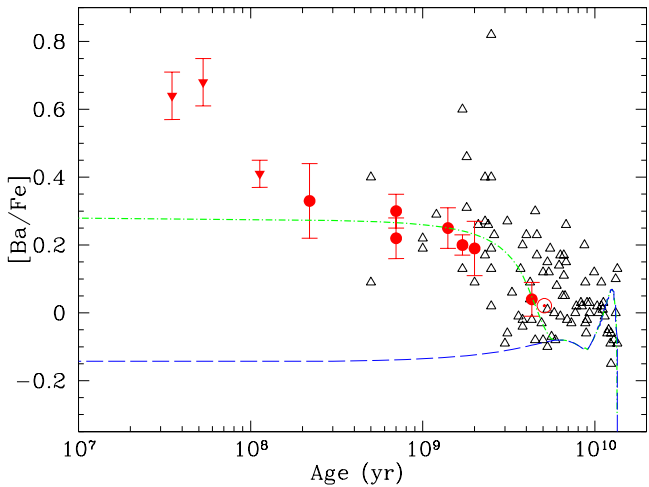}
\caption{Average [Ba/Fe] as function of stellar age for the sub-sample
of clusters whose analysis is based on dwarfs
(filled circles and inverted triangles) compared
with the abundance pattern of disk stars  (open triangles) by
Bensby et al.~\cite{bensby05}. Filled triangles represent
abundance measurements that probably need NLTE corrections. The model
results are shown for two set of yields: ({\em a}\/) standard yields
(Travaglio et al.~1999; Busso et al.~2001), long-dashed curve;
and ({\em b}\/) enhanced $s$-process yields, dot-dashed curve.
Both models show a peak at old ages
due to the $r$-process from massive stars.}
\label{model}
\end{figure}

\section{Time evolution of Ba}
\label{sec:evol}

In Figure~2 we plot the average [Ba/Fe] of OCs (only those based
on dwarf star analysis) and of field
stars by Bensby et al.~\cite{bensby05} vs. stellar age.  
Although field stars
show a larger scatter, the two patterns agree very well and our OCs do
extend the same trend to younger ages. 
Qualitatively,
the observed trend of Ba vs.  age confirms the dominant role played, in
Galactic disk stars, by the $s$-process occurring in low-mass red
giants during their asymptotic giant branch (AGB) phase (Busso et
al.~2001; Cescutti et al.~2006). For a more quantitative comparison, we
also show in Fig.~2 the time evolution of [Ba/Fe] computed with a
Galactic chemical evolution model for the Milky Way. The model adopted
is a generalization of the multi-phase model by Ferrini et
al.~\cite{ferrini92}, originally built for the solar neighborhood, and
later extended to model the entire Galaxy (Ferrini et al.~1994).  We
refer to Magrini et al. (2007, 2008) for a detailed description of
the model.  In our ``standard" model we adopted the average $s$-process
and $r$-process yields by Busso et al.~\cite{busso01} (extrapolated as
in Cescutti et al. 2006) and Travaglio et al.~\cite{travaglio01},
respectively.  While the $r$-process yields do not affect the [Ba/Fe]
ratio in the metallicity range of our OCs (the $r$-process is important
only for $[\mbox{Fe/H}]\lesssim -2.2$), the values of the $s$-process
yields are crucial to reproduce the increasing trend of [Ba/Fe] with
decreasing age of the cluster.

The figure clearly shows that the model with standard yields does not
reproduce the empirical distribution, and, in particular
the observational evidence
that at recent epochs Ba is sensitively enhanced both for field stars
and for OCs. The standard $s$-process yields do not even
fit the solar [Ba/Fe] value and result in a
Ba abundance ratio $[\mbox{Ba/Fe}] \approx -0.15$ at 1 Gyr, to be compared
with the measured value of $\sim 0.2$. 
In the original work from where
the $s$-process yields were taken (Busso et al. 2001) it was underlined
that at every epoch and for every group of stars there was evidence of
a large spread in their efficiencies. However, a simply randomly
distributed spread seems not to be the case: it would generate from
each given stellar generation a yield close to the average one adopted by
us, and this is clearly inadequate to explain the data. Instead, we
``empirically" verified that the observations, showing a very late
increase of Ba in the Galaxy, would require an efficiency of production
strongly weighted toward  lower masses. Namely, the observations can be
well fitted by increasing the yields of stellar masses between 1 and
1.5 $M_\odot$ by a factor of $\sim 6$ with respect to the extrapolation
obtained from the average yields (roughly corresponding to the case
ST/1.5 in Busso et al. 2001).

The results of a chemical evolution model where we have adopted 
``enhanced $s$-process'' yields are shown in Fig.~2.
In this model the solar Ba abundance can be reproduced, with 
contributions from the $s$- and $r$-process of 84\% and 16\%, respectively.
As mentioned, the model with standard yields
has instead a lower percentage of Ba produced 
by the $s$-process at
the time of formation of the Sun. 
More in general, the model with
enhanced yields provides a recent production of Ba in better agreement with
the observations of OCs older than $\sim 500$~Myr. Both models fail to 
fit younger clusters and, indeed, it would be difficult to imagine a
process capable of creating Ba in the last 500~Myr of Galactic
evolution, unless local enrichment is assumed.  We suggest that the
difference between our model predictions and the measured [Ba/Fe]
ratios for young clusters provides an estimate of the NLTE effects that
affect the Ba determination in these stars.

We finally stress
that the classical plots of [Ba/Fe] against [Fe/H] (see, e.g.,
Travaglio et al. 1999, 2001; Cescutti et al. 2006) suffer for a strong
degeneration in terms of age, since they compress the most recent
period. Conversely, the plot of [Ba/Fe] against age favors the youngest
ages and thus is ideal to approach our problem. We also investigated 
the presence of a possible relationship of [Ba/Fe] with metallicity among our
sample, finding that no significant trend is 
present between these two quantities.

\section{Discussion}
\label{sec:discussion}

Our results show for the first time there is a need for an enhanced Ba
production in low-mass stars, mainly from 1 to 1.5~$M_\odot$, formed
from $\sim 10$ to $\sim 5$~Gyr ago (e.g. Charbonnel et al.~1996).  We
have shown that, when adopting the average yields that have been 
commonly used so far, Galactic
evolutionary models are in sharp disagreement with  data. An enhanced
Ba production, like the one assumed in the previous section, seems to
be necessary, but one has to verify whether this ad-hoc data-fitting 
is physically acceptable.

The most promising possibility to obtain enhancements in the Ba
production, especially in low masses, is that the intrinsic spread
found normally in the $s$-process efficiency at any metallicity be not
really a spread, but rather a dependence of the efficiency itself on
the initial mass, in the sense of higher production for lower masses.
As time passes, lower mass stars with larger Ba production start to
contribute, and the average yield changes in time.  Specifically, the
formation of the neutron source $^{13}$C requires a penetration of
protons below the formal convective border at dredge-up. This is a
typical extra-mixing process, like those necessary to account for the
carbon and oxygen isotopic mix in low-mass red giants, and for the
production and destruction of Li in them. These mechanisms are known to
be increasingly more efficient for decreasing stellar mass.  The
evidence coming from the Ba enhancement indirectly suggests that
the mixing processes driving the formation of the neutron source behave
similarly, and are maybe originated by the same cause (e.g. Busso et al.
2007). In general, we verified that the efficiency in $s$-processing
required in very low mass stars in order to explain the data is close
to the upper limit (case ST*2) suggested by Busso et al (2001), not to
the average of the spread there considered (case ST/1.5). This must
come together with the (already known) increase in the dredge-up efficiency 
at low metallicity, so that this large amount of extra Ba can be carried 
to the surface and ejected into the ISM.

Other possibilities were also considered such as:  {\em i}) an increase
in the initial $s$-element abundances of the contributing stars, due to
previous enrichment by mass transfer from more massive companions; and 
ii) an increase in the $r$-process seeds on which slow n-captures occur,
assuming that old disk stars were still characterized by high
$r$-process abundances.  Both attempts proved incapable of increasing 
substantially the Ba yields.

\section{Conclusions}

High-resolution observations of OCs allowed us to derive Ba abundances
in a uniform way in systems covering a wide age range.  The [Ba/Fe]
derived in OCs (but also in field stars, by other authors) shows a
steady increase with time. 
In the framework of Galactic chemical evolution,  this can be
explained by an enhanced $s$-process production of Ba in very old AGB
stars below 1.5 $M_{\odot}$.

This need can be reconciled with the present understanding of the $s$-process if an
anti-correlation exists between the initial mass of the star and the effectiveness of
the extra-mixing mechanism responsible for forming the neutron source.
Stars at the lowest mass limit for contributing would in this case have
a higher efficiency in $s$-processing. Accounting for the observations then 
requires that this efficiency
be close to the upper limit indicated by Busso et al.~(2001), i.e. an
amount of $^{13}$C burnt per interpulse higher by a factor of three 
than the gross average (case ST/1.5) discussed in that work: this would
produce an increase in the Ba yield by a factor of 5-6. This scenario is plausible if
the mechanisms driving the formation of the $^{13}$C pocket are similar
to those promoting the slow circulations known to occur below the convective envelope.

The youngest clusters of our sample, with ages $<500$~Myr, have even
higher [Ba/Fe] ratios. We suggest that this is due to NLTE effects,
and, in particular, to the effects of a strong, warm flux from a
chromosphere. The comparison with our model suggests that NLTE effects
in young stars might increase the [Ba/Fe] ratios by a factor of two of
more. Independent estimates of these effects from NLTE calculations as
well as a more complete analysis of the hypothesis here advanced of an
anticorrelation between initial mass and $s$-process efficiency are
strongly needed.

\begin{table*}
\begin{center}
\caption{OC sample}
\begin{tabular}{lrrrrrr}
\tableline
          &            &         &          &          &          &                  \\
 Cluster  &  Age (Gyr) &  [Fe/H] &  Ref.    &  [Ba/Fe] & \# stars & $T_{\rm eff}$ (K)\\
          &            &         &          &          &          &                  \\
 IC 2602           & 0.035 &  $0\pm 0.01$      & 1 & $0.64\pm 0.07$ & 8 & 4770--5760 \\
 IC 2391           & 0.053 &  $-0.01\pm 0.02$  & 1 & $0.68\pm 0.07$ & 6 & 4680--5970 \\
 NGC 2516          & 0.110 &  $0.01\pm 0.07$\tablenotemark{a} & 2 & $0.41\pm 0.04$  & 4 & 5110--5659\\
 NGC 6475          & 0.22  &  $0.14\pm 0.06$   & 3 & $0.33\pm 0.11$  & 19  & 5048--5888\\
 Hyades            & 0.7   &  $0.13\pm 0.05$\tablenotemark{b} & 2 & $0.30\pm 0.05$  & 4 & 5079--5339\\
 Praesepe          & 0.7   &  $0.27\pm 0.04$   & 4 & $0.22\pm 0.06$  & 6 & 5720--6280\\
 NGC 3680          & 1.4   &  $-0.04\pm 0.03$  & 4 & $0.25\pm 0.06$  & 2 & 6010--6210\\
 IC 4651           & 1.7   &  $0.12\pm 0.05$   & 4 & $0.20\pm 0.03$  & 5 & 5910--6320 \\
 NGC 752           & 2.0   &  $0.01\pm 0.04$   & 5 & $0.19\pm 0.08$  & 9 & 5531--6151 \\
 M67                   & 4.3   &  $0.02\pm 0.14$   & 6 & $0.04\pm 0.05$ & 10  & 5541--6223 \\
                   &  &                      &      &  & & \\
\tableline
           &                     &      &      &   &          &   \\
NGC 2324        &  0.7 & $-0.17\pm 0.05$   & 7 & $0.66\pm 0.09$ &7 & 4300--5100 \\
NGC 3960        &  0.7 & $0.02\pm 0.04$    & 8 & $0.50\pm 0.06$ &6 & 4870--5050\\
NGC 2660        &  0.8 & $0.04\pm 0.04$    & 8 & $0.47\pm 0.06$ &5 & 5030--5200\\
NGC 2477        &  1.0 & $0.07\pm 0.00$    & 7 & $0.40\pm 0.03$ &6 & 4950--5030\\
Berkeley 20     &  4.3 & $-0.30\pm 0.02$   &10 & $0.09\pm 0.20$ &2 & 4400--4770  \\
Berkeley 29     &  4.3 & $-0.31\pm 0.03$   &10 & $0.40\pm 0.10$ &5 & 4970--5020\\
NGC 6253        &  4.5 & $0.39\pm 0.08$    &9  & $0.14\pm 0.10$ &5 & 4450--6200\\
Melotte 66       &  5.4 & $-0.33\pm 0.03$   &10 & $0.32\pm 0.10$ &6 & 4717--4850\\
Berkeley 32     &  6.1 & $-0.29\pm 0.04$   &8  & $0.24\pm 0.15$ &9 & 4760--4920 \\
Collinder 261        &  8.4 & $0.13\pm 0.05$    &10 & $0.22\pm 0.03$ &7 & 4350--4720  \\
                 &     &                      &     &  &        & \\

\tableline
\end{tabular}
\tablecomments{Results of [Ba/Fe] ratios for our sample OCs.  We list:
ages (Col.~2) from Magrini et al.~(2008), with the exception of
NGC~2516 (WEBDA) and NGC~752 (Sestito et al. 2004); [Fe/H] values along
with references for them and stellar parameters (Cols. 3 and 4);
average [Ba/Fe] ratios along with the standard deviation (Col. 5);
number of stars and range of effective temperatures (Cols. 6 and 7).
Reference codes are as follows:  1=D'Orazi \& Randich (2008), 2=Randich
et al. (2007), 3=Sestito et al.  (2003), 4=Pace et al. (2008),
5=Sestito et al. (2004),6=Randich et al.  (2006), 7=Bragaglia et al.
(2008), 8=Sestito et al. (2006), 9=Sestito et al. (2007), 10=Sestito et
al. (2008).}
\label{results}
\tablenotetext{a}{Terndrup et al. (2002)} \tablenotetext{b}{Paulson et
al. (2003)}
\end{center}
\end{table*}

\acknowledgments

We are grateful to A.~McWilliam and C.~Sneden for many helpful
suggestions.  We warmly thank L.~Belluzzi and G.~Cauzzi for useful
discussions. The anonymous referee is acknowledged for a careful reading of the paper and valuable comments. 
This work has made use of the WEBDA database, originally
developed by J.~C.~Mermilliod, and now maintained by E.~Paunzen.

{}

\end{document}